\documentclass[english,aps,prb,floatfix,superscriptaddress,amsmath,amssymb,notitlepage]{article}
\usepackage[T1]{fontenc}
\usepackage[utf8]{inputenc}
\usepackage{geometry}
\usepackage{float}
\usepackage{amsbsy}
\usepackage{xcolor}
\usepackage{amssymb}
\usepackage{amsmath}
\usepackage{graphicx}
\usepackage{babel}
\usepackage{graphicx}
\usepackage{caption}
\usepackage{subcaption}
\usepackage{hyperref}
\usepackage{xcolor}
\usepackage{authblk}
\usepackage[bibstyle=phys,citestyle=numeric,sorting=none,isbn=false,url=false,doi=false,eprint=false,maxbibnames=5,biblabel=brackets,articletitle=false,chaptertitle=false]{biblatex}
\begin{document}
\def\newblock{}
\title{Magnetic ordering in VI$_3$: a van der Waals material combining vastly different magnetic anisotropies}

\author[1,*]{K. K. Pokhrel}
\author[1]{S. Ray}
\author[1]{N. Mach\'a\v{c}ov\'a}
\author[1]{A. Koliogiorgos}
\author[1,*]{K. Carva}

\affil[1]{Charles University, Department of Condensed Matter Physics, Ke Karlovu 5, 121 16 Prague 2, Czech Republic}

\affil[*]{Corresponding authors: krishna-kumar.pokhrel@matfyz.cuni.cz, karel.carva@matfyz.cuni.cz}
\date{}

\maketitle

\begin{abstract}
Among magnetic van der Waals materials, the vanadium trihalide family exhibits unique features. In particular, \(\mathrm{VI}_3\) can contain two types of V atoms because two electronic occupations are energetically close and may coexist in real samples. These types have strikingly different magnetic anisotropies, predicted to differ by more than an order of magnitude. \(\mathrm{VI}_3\) also shows an unusual thickness dependence: the reported monolayer Curie temperature, \(T_{\mathrm C}\), is higher than that of the bulk, contrary to the usual expectation that interlayer coupling reinforces magnetic order.

Here, we use atomistic spin-dynamics simulations to study critical temperatures in \(\mathrm{VI}_3\) from the combined perspective of single-ion anisotropy and exchange interactions, aiming to identify the microscopic origin of the bulk--monolayer anomaly. We consider a system composed of two V types with properties predicted by first-principles calculations. The anisotropy contrast strongly affects thermal stability: increasing the fraction of high-anisotropy sites raises the energy cost of transverse spin fluctuations and increases \(T_{\mathrm C}\). At the same time, the interlayer super-superexchange network is modified by the inhomogeneous V environment. It becomes spatially non-uniform and can contain competing exchange pathways, weakening coherent interlayer magnetic order while preserving robust intralayer ferromagnetic correlations.

Our results show that changing the ratio of the two V types can cause a substantial shift in \(T_{\mathrm C}\). The experimental bulk value is reproduced when the ratio is close to 1:1, consistent with two experimental indications of coexisting V configurations. Thus, the finite-temperature magnetization behavior provides further support for the coexistence of two V configurations in \(\mathrm{VI}_3\). The experimentally observed monolayer \(T_{\mathrm C}\) is obtained for a slightly modified ratio, which may be related to the polaron concentration. The sensitivity of \(T_{\mathrm C}\) to the HO/LO ratio suggests that the ordering temperature of \(\mathrm{VI}_3\) could be tuned over a broad range by controlling the relative occupation of the two vanadium configurations.
\end{abstract}
\noindent\textbf{Keywords:} 2D vdW materials, atomistic spin dynamics, orbital configurations, magnetic anisotropy

\section{Introduction}

The weak interlayer bonding of van der Waals (vdW) materials allows their layers to be isolated, re-stacked, and arranged with controlled sequence and stacking geometry, providing access to 2D physics with physical properties that are often very different from their bulk counterparts  \cite{Geim2013,Ajayan2016_Review_2D}.  Dimensionality also plays an essential role in magnetic properties \cite{Burch2018_2DmReview_Phenomena,Gibertini2019_Magn2D_NatNano_Review}. The magnetic interlayer coupling is much weaker than its intralayer counterpart here, giving rise to the quasi-2D character of magnetism. Intralayer couplings are generally much more robust against perturbation than the interlayer ones. vdW engineering methods allow to change relative arrangement of layers in terms of their relative rotation (twisting) or stacking order. In these cases, the geometry of individual layers is not significantly changed, but the different geometry of interaction paths between atoms of different layers will strongly affect the magnetic interactions between layers \cite{Jiang2018_ElFieldMag_CrI3}.  This volatility may allow us to tune it within a wide range of values. For example, in CrI$_3$ even sign of the interlayer magnetic exchange changes when the system is thin enough, so that in a bilayer individual layers are antiferromagnetically coupled \cite{Huang2017_FM_2D_CrI3}. Due to the weak vdW coupling between layers, the stacking order of layers is most easily affected. Different stackings lead to significant changes in interlayer coupling \cite{Sivadas2018_Stacking_Mag_CrI3}. Pressure experiments are expected to affect mainly the vdW gap size and thus again the interlayer coupling. The pressure effect can be further enhanced if it leads to stacking change, different stacking has been seen to again change the sign of interlayer interaction in CrI$_{3}$ \cite{Song2019_Switch2DMag_Pressure_CrI3,Li2019_Pressure_Stacking_IntLExch},  CrBr$_{3}$ \cite{Misek2025_CrBr3}.
or VI$_{3}$ \cite{Gordon2021_MagInt_VI3}. Interlayer coupling has also been shown to be tunable by electron/hole doping \cite{Jiang2018_ElFieldMag_CrI3}. Note that the regime with interlayer interaction much weaker than the intralayer one represents a significant challenge for the spin-wave theory \cite{Irkhin1999_Katsn_LayMagnetsThe}.

In a magnetic monolayer (ML), the interlayer coupling is of course absent; hence, it can be seen as the limiting case of a layered vdW magnet. Due to the Mermin-Wagner theorem such truly 2D systems feature long-range magnetic order at sufficiently low temperatures only if magnetic anisotropy is present  \cite{Burch2018_2DmReview_Phenomena,Gibertini2019_Magn2D_NatNano_Review}, but their critical temperature is typically reduced with respect to the bulk (quasi-2D) case. Although the interlayer coupling is weak, it still acts in favor of magnetic order. Its absence should therefore destabilize the magnetic order and reduce the critical temperature $(T_{\mathrm{C}})$. 
Comparison of bulk and monolayer behavior across vdW trihalides MX$_3$ shows that reducing the thickness does not lead to a universal evolution of magnetic order, but instead exposes how strongly the ordered state depends on interlayer coupling and magnetic anisotropy. In CrI$_3$, bulk ferromagnetism persists up to $T_{\mathrm C}\approx 61$ K, whereas monolayer ferromagnetism survives in the 2D limit with reduced $T_{\mathrm C}\approx45$ K\cite{Huang2017}. The bilayer already switches to an interlayer-antiferromagnetic state that requires a field of about $0.65$ T to recover ferromagnetic alignment, demonstrating the strong stacking sensitivity of interlayer exchange \cite{Sivadas2018_Stacking_Mag_CrI3,Gibertini2019}.
 CrBr$_3$ behaves more conventionally, retaining ferromagnetism down to the monolayer limit with only a modest reduction in $T_{\mathrm C}$ relative to bulk, consistent with predominantly stabilizing interlayer coupling in the usual stacking \cite{Zhang2019,Gibertini2019}. CrCl$_3$ provides a complementary case of easy-plane antiferromagnet, showing that anisotropy can redirect thin-limit behavior even when the magnetic order remains robust within individual layers \cite{Cai2019,Gibertini2019}. On the other hand, CrGeTe$_3$ shows a much stronger suppression of magnetic order upon thinning, with the bilayer exhibiting an effective transition temperature of only about one-half of the bulk value of 61 K, while Fe$_3$GeTe$_2$ remains magnetic down to the monolayer limit but with a pronounced thickness dependence of $T_{\mathrm C}$, and FePS$_3$ retains antiferromagnetic order even in atomically thin layers \cite{Gibertini2019}.

Interestingly, an exception to this observation has recently been found in VI$_{3}$, where 1ML $T_{\mathrm{C}}$ reaches 60 K \cite{Lin2021_VI3_MCD_ML} compared to the bulk $T_{\mathrm{C}}$ of $50$ K \cite{Son2019_Bulk_VI3}. 
An analogous bulk--monolayer enhancement has not yet been established as a general feature of the related VBr$_3$ and VCl$_3$ compounds, although their magnetic properties are also strongly sensitive to orbital and structural degrees of freedom \cite{Hovancik2023_VBr3_IntralayAFM,Gu2024_VBr3,Deng2025_VCl3_ML_FEl}.
This finding is in striking contrast to a recent study based on spin-wave theory on VI$_{3}$, which correctly predicts $T_{\mathrm{C}}$ for the bulk but reduces it to $\sim14$ K for 1ML \cite{Gordon2021_MagInt_VI3}. This calls for a deeper theoretical investigation of finite-temperature magnetism in VI$_3$ and the role of interlayer coupling. An enhancement of magnetism with reduced dimensions has been observed. Still, it is typically connected with an associated change of electronic structure connected to intralayer magnetic properties, for example, the flattening of bands so that the Stoner criterion may become satisfied in a thin layer rather than bulk. However, in vdW materials, the influence of different layers on intralayer properties is limited.

A deeper understanding of magnetic properties requires knowledge of the electronic structure of the VI$_{3}$ semiconductor. In VI$_{3}$ as well as other transition metal trihalides, the local environment around the magnetic atom is octahedral with a small trigonal distortion, and electronic states in the $d$-shell can thus be expected to split initially as expected for the $O_{h}$ group. The $t_{2g}$ shell of V is only partially occupied, and the trigonal field further splits it so that the ${e'_{g}}^2$ or ${a_{1g}}^1 {e'_{g}}^1$ configurations are occupied. The latter case together with the effect of spin-orbit interaction allows for an exceptionally high value of the orbital angular momentum V \cite{Yang2020_VI3_2DIsing_CFdiag}. The configuration with a high orbital moment (hereafter HO) has been predicted as the ground state by several \textit{ab initio} approaches \cite{Yang2020_VI3_2DIsing_CFdiag,Sandratskii2021_VI3}; however, other calculations found the low-orbital-moment \(e_g'^{\,2}\) configuration (hereafter LO) to be the ground state \cite{Vita2022_OrbChar_VI3_CrI3_GS_ARPES}. HO-type vanadium has been observed experimentally in VI$_3$ samples, while inclusion of an LO-type contribution was required to reproduce the measured x-ray absorption and x-ray magnetic circular dichroism spectra \cite{Hovancik2023_VI3_XMCD_OrbMom}. This coexistence was also indicated in neutron scattering data \cite{Lane2021_SO_excit_VI3,Hovancik2023_VI3_XMCD_OrbMom}.
The preference for either of these solutions depends on a delicate competition between electron correlations, lattice distortions, spin-orbit interaction, and the Kugel-Khomskii orbital ordering mechanism \cite{Solovyev2024_FM_FE_Kugel-Khomskii_Hund_VI3}. Supporting this picture, a recent DFT+U study of VI$_3$ and VCl$_3$ found competing $a_{1g}$- and $e'_g$-based insulating states to lie close in energy (of order 10--50 meV) and showed that symmetry-breaking lattice distortions can strongly reorganize their relative stability and orbital character \cite{Camerano2024}, highlighting the sensitivity of the electronic state to local structural distortions. As HO and LO solutions contain different orbital occupations, it may lead to strikingly different exchange interaction of both intralayer and interlayer types. They also differ in magnetic anisotropy, which in fact reaches exceptionally high values for the high momentum states \cite{Yang2020_VI3_2DIsing_CFdiag,Sandratskii2021_VI3}. 

VI$_{3}$ undergoes small lattice distortions that lead to deviations from the high-temperature $\bar{R}3$ structure \cite{Son2019_Bulk_VI3, Dolezal2019_StructTrans_VdW_VI3,Hovancik2022_THz_IR_Raman_VI3} and is also susceptible to stacking faults \cite{Marchandier2021_LiVI3_Struct}. Furthermore, x-ray photoemission spectroscopy in VI$_{3}$ has shown a significant presence of $V^{2+}$ atoms, which have been argued to form polarons here \cite{Mastrippolito2023_VCl3_Polaronic_MottInt}. The existence of both high- and low momentum states in real samples and the resulting inhomogeneity can, in fact, be ascribed to the local lattice perturbations introduced by polarons. 
 Although basic properties of LO and HO atom types have been proposed, a simulation comparing the finite-temperature behavior of these two types separately as well as in a mixture is missing. Thus, the new contribution of the present work is to connect the established HO/LO coexistence picture to finite-temperature magnetic ordering by showing how anisotropy contrast and spatially non-uniform interlayer exchange affect $T_C$, including the anomalous bulk--monolayer difference in VI$_3$.

In this paper, we use atomistic spin-dynamics simulations to understand the behavior of critical temperatures in this system from the combined perspective of single-ion anisotropy and exchange interactions, with a focus on identifying the microscopic origin of the bulk/monolayer VI$_{3}$ $T_\mathrm{C}$ difference.
First, we study the dependence of $T_\mathrm{C}$ on the magnitude of magnetic anisotropy and the interlayer coupling. Next we consider the model composed of two different types of V  with properties predicted by first principles calculations. Our results show that variation of the ratio between different V types can cause a rather large change in $T_{\mathrm{C}}$. We decompose this change into the effects of anisotropy and exchange modifications. We have predicted what concentrations describe the bulk and ML critical temperatures, and how the difference between them changes with V type concentration.

\section{Theory and methods}
\subsection{Layered spin Hamiltonian}

\begin{figure}[H]
\centering
\includegraphics[width=0.46\linewidth]{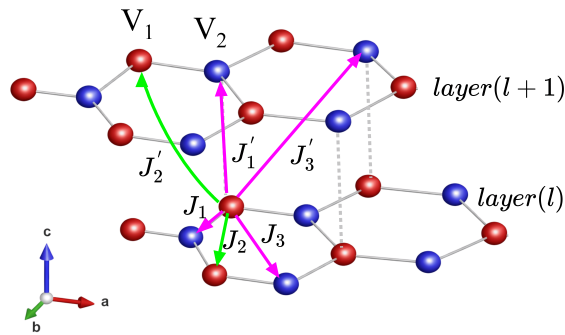}
\caption{Schematic illustration of the exchange pathways in
\(\mathrm{VI}_3\). Intralayer interactions are denoted by \(J_i\), while
the interlayer interactions connecting layer \(l\) to layer \(l+1\) are
denoted by \(J'_i\). \(J_1\), \(J_2\), and \(J_3\)
correspond to the first-, second-, and third-nearest-neighbor
intralayer interactions.  \(J'_1\), \(J'_2\), and \(J'_3\) 
represent the first-, second-, and third-nearest-neighbor interlayer interaction. Red and blue spheres distinguish the two inequivalent V
sites shown in the schematic.}
\label{fig:exchange_scheme}
\end{figure}

The magnetic properties of bulk VI$_3$ are mapped to a
layered classical Heisenberg spin Hamiltonian. Because VI$_3$ is a van der Waals
magnet, the magnetic interactions can naturally be separated into
strong intralayer exchange interactions between V atoms within the same
honeycomb layer and considerably weaker interlayer interactions between
adjacent layers. The interaction network considered in the present work
is illustrated schematically in Fig.~\ref{fig:exchange_scheme}.

The interaction terms in the Heisenberg Hamiltonian can be
explicitly separated into intralayer exchange, interlayer exchange, and
single-ion anisotropy contributions. Spin orientations are described by  \(S_{l,i}\) with a unique layer index \(l\) and magnetic site index  \(i\). In a quasi-2D system, boundaries may be neglected and  the exchange interactions are equivalent for all layers due to the translational invariance along the \textit{c} axis. Within each honeycomb layer, the magnetic interactions are then represented by the
intralayer exchange \(J_{ij}\). Adjacent layers are coupled through the
interlayer exchange interactions \(J'_{ij}\) and \(J''_{ij}\), where sites from layer \(l\)  are connected by \(J'_{ij}\)  to those from adjacent layer \(l+1\), and by  \(J''_{ij}\)  to those from layer \(l-1\). Because the interlayer exchange is much weaker than the intralayer exchange, interactions between non-neighboring layers are neglected. The  single-ion anisotropy in the Hamiltonian is of uniaxial type with site-dependent strength \(K_i\). With these assumptions the Hamiltonian is given as

\begin{equation}
H = H_{\parallel}+H_{\perp}+H_{\mathrm{SIA}}
= -\sum_l\sum_{i,j}J_{ij}S_{l,i}S_{l,j}
-\sum_l\sum_{i,j}
\left(
J'_{ij}S_{l,i}S_{l+1,j}
+
J''_{ij}S_{l,i}S_{l-1,j}
\right)
-\sum_l\sum_i K_i\left(S^z_{l,i}\right)^2,
\label{hamiltonian}
\end{equation}

We adopt the standard notation of atomistic spin dynamics, in
which the classical spins \(S_{l,i}\) are normalized to unity
\cite{Evans2014_ReviewASD}. Heisenberg Hamiltonian is traditionally defined with a summation that includes each exchange coupling either once (summation over $i<j$) or twice (summation over $i,j$). We employ here the latter choice, which is more common in atomistic spin dynamics\cite{Evans2014_ReviewASD}. The interlayer interaction is defined in the same manner, so that the coupling in both directions is present in the Hamiltonian. Note that in order to describe intralayer interactions without the need for layer specification one needs to have a reasonably well defined relation between \(i\) and \(j\) from consecutive layers, and interlayer directions have to be distinguished by the notation.
%- \(J'_{ij}\) or \(J''_{ij}\).  
%Thus, the same physical bond is represented as \(J'_{ij}\) from one endpoint and as \(J''_{ji}\) from the other.
Then employing the translational invariance, \(J''_{ij}=J'_{ji}\).

Interlayer interactions are generally small, but there are a number of different interactions with a comparable magnitude (some of these are shown in Fig.~\ref{fig:exchange_scheme}). It is not clear for how big distances these interactions can be neglected. Furthermore, the neighbor distribution is different for different sublattices. Presentation of few individual values (as \(J'_1\) - \(J'_3\) ) is thus insufficient for understanding the interlayer relations. Therefore the best characterization of interlayer interaction is provided by their sum over all sites in one simulation cell layer, divided by the total number of such sites \(N\);  we denote its value  as the effective total interlayer interaction \(J_L\):

\begin{equation}
J_{L}=\frac{1}{N}\sum_{i,j}
\left(J'_{ij}+J''_{ij}\right),
\label{eq:J_L-def}
\end{equation}

This quantity can be obtained from the energy difference between the ferromagnetic (FM)  and layer-antiferromagnetic (LAFM) configurations  \cite{Sivadas2018_Stacking_Mag_CrI3}.
In the LAFM state, each layer remains internally ferromagnetic, while adjacent layers have
opposite magnetization directions. Then we assume that for even layers ($l=2p$) $S_{2p,i}=1$, while for odd ones ($l=2p+1$) $S_{2p+1,i}=-1$.  
The intralayer and single-ion-anisotropy contributions do not change when switching between the FM and LAFM configurations. Then the difference between these two configurations is for a simulation cell with \(N_L\) layers given as

\begin{equation}
E_{\mathrm{LAFM}}-E_{\mathrm{FM}}=-\sum_{l}\sum_{i,j}(J'_{ij}+J''_{ij})\left(-1-1\right)=2N_{L}NJ_{L}\ .
\label{eq:E_LAFM}
\end{equation}

\subsection{Atomistic spin dynamics}

Finite-temperature magnetic properties were calculated using the Uppsala Atomistic Spin Dynamics (UppASD) framework, where exchange and anisotropy parameters are mapped onto an atomistic spin Hamiltonian \cite{UppASD,Eriksson2017,Skubic2008}. The simulations were performed for a hexagonal \(30\times30\times10\) supercell with periodic boundary conditions in all directions. Within the adiabatic approximation, the local magnetic moments are treated as classical vectors whose time evolution follows the stochastic Landau--Lifshitz--Gilbert equation. Thermal fluctuations were introduced through a Gaussian white-noise field satisfying the fluctuation--dissipation theorem \cite{Eriksson2017}. At each temperature, the system was equilibrated for \(90\,000\) Monte Carlo steps, after which the magnetization was averaged over an additional \(90\,000\) spin-dynamics steps using a time step of \(\Delta t=10^{-16}~\mathrm{s}\). For each value of \(C_{\mathrm{HO}}\), the results were averaged over \(M_{\mathrm{ensemble}}=5\) independent ensemble runs, each generated with a random HO/LO assignment on the V sublattice. The spread among the corresponding values $T_{\mathrm{C}}$ was used to estimate the statistical uncertainty.

To implement the effective interlayer coupling, the average quantity \(J_L\) in Eq.~\ref{eq:J_L-def} was converted into pairwise interlayer interactions \(J'_{ij}\) between adjacent layers. In the homogeneous single-environment model, \(J_L\) was distributed uniformly over all V--V pairs within the cutoff distance \(r_c=2.0817~\text{\AA}\), so that the sum of the included bonds reproduced the prescribed value of \(J_L\). Because adjacent hexagonal layers are laterally displaced, the number of interlayer neighbors within the cutoff is not identical for every site. This variation is purely geometrical and does not introduce additional magnetic inhomogeneity.

In the two-environment model, the same lattice geometry was retained, but the local spin-Hamiltonian parameters were allowed to depend on the V environment. Two inequivalent environments, denoted HO and LO, were introduced to represent local orbital or structural configurations with different anisotropy, intralayer exchange, and interlayer exchange parameters. Thus, the values of \(K\), \(J_1\), and \(J_L\) were assigned according to the corresponding local V environment, as listed in Table~\ref{heisenberg_parameters}. For mixed HO–LO bonds, the exchange parameter is taken as the arithmetic mean of the corresponding HO–HO and LO–LO values. This approximation is motivated by the orbital-dependent superexchange picture discussed by Yang \textit{et al.}\cite{Yang2020_VI3_2DIsing_CFdiag}. This picture rests on the relatively stronger hopping between ${e'_{g}}$ orbitals (see Eq. (2) and Fig. 3 of \cite{Yang2020_VI3_2DIsing_CFdiag}). This hopping enhances the FM interaction for the ${a_{1g}}^1{e'_{g}}^1$ - ${a_{1g}}^1{e'_{g}}^1$ occupied pair (HO-HO case), where it contributes twice, while it does not contribute for the ${e'_{g}}^2$-${e'_{g}}^2$ pair (LO-LO) case. For the ${a_{1g}}^1{e'_{g}}^1$ - ${e'_{g}}^2$ occupied pair (HO-LO) it contributes once, e.g. by half the difference between HO-HO and LO-LO. The model should therefore be understood as an effective description of orbital-driven local inhomogeneity on the V sublattice, rather than as random chemical substitution.

The Curie temperature \(T_C\) was determined from the temperature dependence of the magnetic susceptibility. Although the magnetic transition can also be identified from other quantities, such as heat capacity or Binder cumulant, we use the peak position of the susceptibility as a practical criterion for \(T_C\). This criterion is appropriate because the susceptibility shows a pronounced maximum at the transition, coinciding with a rapid loss of magnetization \(M\).

\subsection{First principles calculations}
Density functional theory (DFT) calculations employed the full-potential linear augmented plane wave (FP-LAPW) method, as implemented in the band structure program ELK \cite{elk_code}. The generalized gradient approximation (GGA) parametrized by Perdew-Burke-Ernzerhof \cite{r_96_Perdew_GGAsimple} has been used to determine the exchange-correlation potential. Spin-orbit coupling (SOC) is known to play a key role in V trihalides and has been included in the calculation \cite{Yang2020_VI3_2DIsing_CFdiag, Sandratskii2021_VI3}. We have included the effect of electron-electron correlations in terms of the Hubbard correction term $U= 4.3$ eV and Hund exchange $J_H= 1.1$ eV \cite{Anisimov1991_LDAU, Anisimov1997_LDAU_review} acting on 3d electrons of V. Double counting was treated in the fully localized limit. The entire Brillouin zone has been sampled by $10 \times 10 \times 5$ k-points and the convergence with respect to k-mesh density has been verified. To evaluate interlayer interaction $J_{L}$ we have used unit cell doubled in the z direction. The energies of the self-consistent ground states were then calculated with forced ferromagnetic (FM) and antiferromagnetic (AFM) alignment between layers.
HO and LO configurations differ in the orbital occupation of the trigonally split $t_{2g}$ manifold. Because these two occupations are nearly degenerate but coupled to slightly different local lattice distortions (a Jahn-Teller/Kugel-Khomskii-type effect), we bias the calculation toward one solution or the other by starting from a slightly modified trigonal distortion associated with each configuration, and let the calculation reach self-consistency with the same DFT+$U$+SOC Hamiltonian; this is the same procedure used in our earlier VBr$_3$ study. \cite{Hovancik2023_VBr3_IntralayAFM}. This particular distortion can be expressed in terms of the planar distance
(denoted $r$) of the anions from the hollow site in the honeycomb lattice, while
their $z$ coordinate is fixed; see Fig.~6(a) of \cite{Hovancik2023_VBr3_IntralayAFM}.
A slight change of this parameter by $0.01a$ is sufficient to achieve
self-consistently either the HO or LO solution. A related sensitivity of the electronic state to lattice distortions has recently been investigated by Camerano and Profeta \cite{Camerano2024}. They showed that a sufficiently large displacement along an optical $\Gamma$-point phonon eigenvector can drive the system onto a different electronic energy surface and stabilize a symmetry-broken $a_{1g}$-occupied insulating state.

\section{Results}
The predicted values of key Heisenberg Hamiltonian parameters, as shown in Table \ref{heisenberg_parameters}, differ significantly between the two types of V configurations. We first ask how their variation influences the ordering temperature and what can be said about it in general. This is a problem that can be studied on a homogeneous system where we selectively vary its selected parameters.  The effect of the intralayer interaction on $T_{\mathrm{C}}$ is relatively clear and easy to study within simple models; one can expect a linear dependence of  $T_{\mathrm{C}}$. Therefore, in the following subsections we study the influence of $K$ and $J_L$  in more detail.

\subsection{The effect of single-ion anisotropy}

The single-ion uniaxial anisotropy (SIA) for VI$_3$ in the low-orbital-momentum state with the electronic configuration $e'{_{g}}^2$ has been reported to reach values below 0.4 meV \cite{Wang2020_VI3_magFPcalc,Gordon2021_MagInt_VI3,Subhan2020_FPCalc_MAE_VI3} . In the \(a_{1g}^{1}e_g'^{\,1}\) high-orbital-moment configuration, large SIA values of \(15\)--\(20\) meV are predicted by \textit{ab initio} calculations  \cite{Yang2020_VI3_2DIsing_CFdiag,Sandratskii2021_VI3}.  The partially unquenched orbital degree of freedom of V produces a sizable magnetic anisotropy, which is expected to play a central role in stabilizing long-range magnetic order in this layered compound. To isolate the effect of single-ion anisotropy,  $K$ is varied over the range that covers the reported low- and high-orbital-momentum limits, while the exchange parameters are fixed to the HO values listed in Table~\ref{heisenberg_parameters}.

\begin{figure}[H]
\centering
\begin{subfigure}{0.46\textwidth}
\centering
\includegraphics[width=\linewidth]{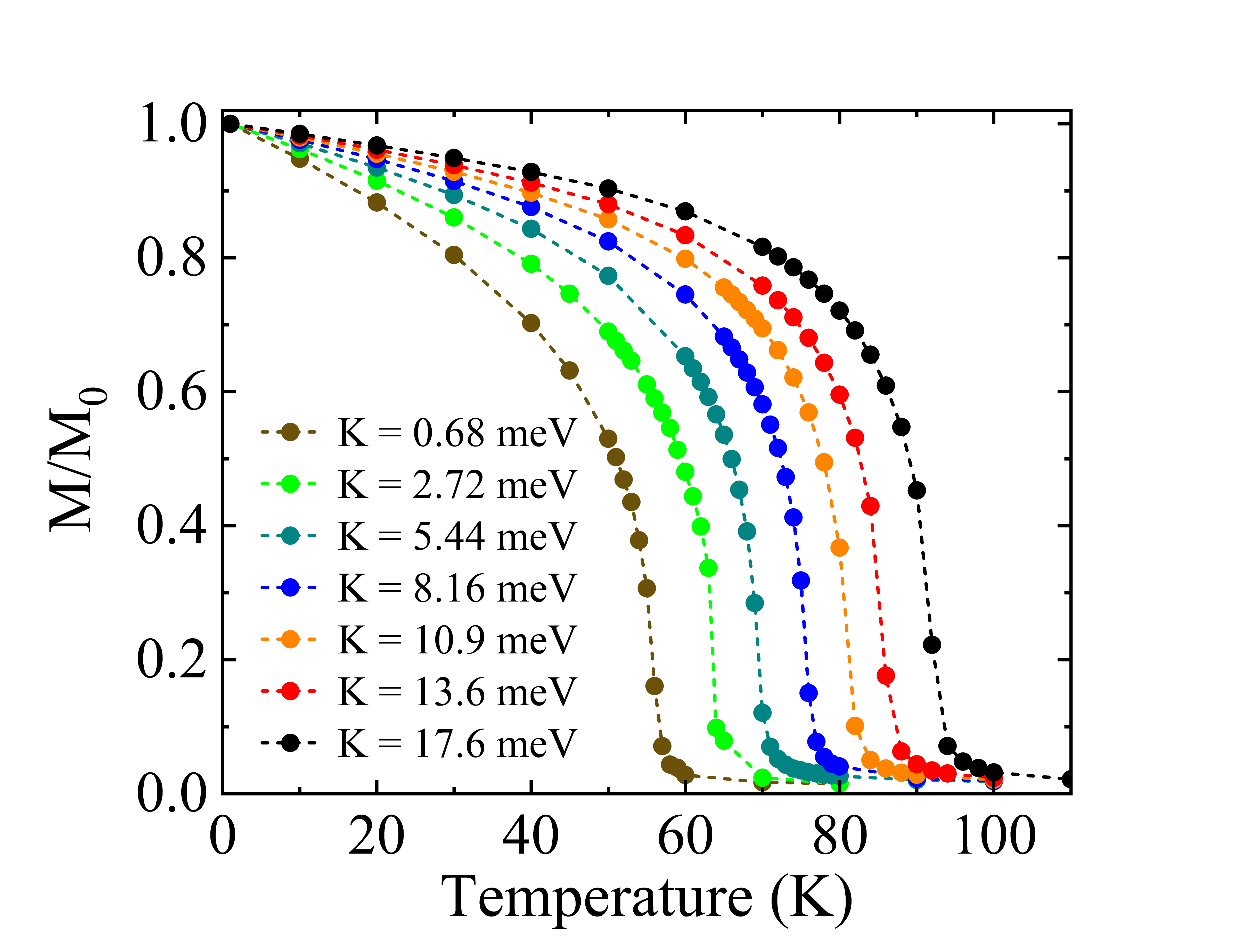}
\caption{}
\label{fig:mag_vs_t_k}
\end{subfigure}
\hfill
\begin{subfigure}{0.46\textwidth}
\centering
\includegraphics[width=\linewidth]{VI3_quasi2D_Vs_TO_bulk.png}
\caption{}
\label{fig:tc_vs_k}
\end{subfigure}
\caption{Effect of single-ion anisotropy in the homogeneous bulk model of \(\mathrm{VI}_3\). (a) Normalized magnetization \(M/M_0\) as a function of temperature for different values of \(K\). (b) Curie temperature \(T_{\mathrm C}\) as a function of \(K\), obtained from UppASD simulations and from the two-dimensional Torelli--Olsen (2D T--O) analytical expression \cite{Torelli2018_CalcTc_2D}.}
\label{fig:anisotropy_dependence}
\end{figure}

Figure~\ref{fig:anisotropy_dependence} shows that the critical temperature increases monotonically with $K$, but the dependence is sub-linear and would tend to saturate at larger anisotropy (above the examined range). Note that for the minimum anisotropy (approaching the Heisenberg limit), the magnetic order is stabilized mainly by the interlayer interaction addressed in the next section.  A stronger out-of-plane anisotropy suppresses transverse spin fluctuations, opens a magnon gap, and therefore stabilizes the ferromagnetic state against thermal disorder. Single-ion anisotropy increases the energy cost of spin deviations and thus controls the thermal stability of the ordered phase (enhances $T_{\mathrm{C}}$). The effect is particularly important in layered magnets, where reduced dimensionality makes the ordered phase more sensitive to thermal fluctuations.
The two possible configurations of V correspond to the lower and upper boundary of the SIA range shown in Fig. \ref{fig:anisotropy_dependence} , and the difference between them can therefore lead to a change of  $T_{\mathrm{C}}$  by almost a factor of 2.

Torelli and Olsen have studied magnetic order stability for a two-dimensional anisotropic Heisenberg ferromagnet with nearest neighbor interaction $J_1$ only \cite{Torelli2018_CalcTc_2D}. They have examined how the critical temperature differs from that obtained in the limit of infinite anisotropy (Ising model) , which is given by \cite{DIXON2005} 
\begin{equation}
T_{\mathrm C}^{\mathrm{Ising}}=
\frac{2J_1\tilde{T}_{\mathrm C}}{k_B}.
\label{eq:ising_tc}
\end{equation}
when converted to our Hamiltonian notation. Here $\tilde{T}_{\mathrm{C}}=1.52$ is the dimensionless Ising critical temperature  for the honeycomb lattice. They have proposed  the following analytical expression for the Curie temperature as a function of the ratio between anisotropy and exchange by fitting classical Monte Carlo simulation results.
\begin{equation}
T_{\mathrm C}=T_{\mathrm C}^{\mathrm{Ising}}f\left(\frac{K}{2J_1}\right),
\qquad
f(x)=\tanh^{1/4}\left[
\frac{6}{N_{\mathrm{nn}}}\ln(1+\gamma x)
\right],
\label{eq:torelli_function}
\end{equation}

 For the honeycomb lattice, \(N_{\mathrm{nn}}=3\) is the number of nearest neighbors of each V site, \(\gamma=0.033\), and \(T_{\mathrm C}^{\mathrm{Ising}}=114\) K. The resulting dependence of \(T_{\mathrm C}\) on \(K\) is shown together with the UppASD results in Fig.~\ref{fig:anisotropy_dependence}(b). The analytical formula apparently captures the change of $T_{\mathrm{C}}$ with reasonable precision, the largest difference to our simulation of quasi-2D system occurring for a low anisotropy. The two graphs can be seen as vertically shifted with respect to each other; this can be ascribed to the interlayer interaction missing in the pure 2D model. We focus on this property in the next subsection. Nevertheless, the change of $T_{\mathrm{C}}$ due to the change of $K$ appears to be predicted by Eq. (\ref{eq:torelli_function}) with accuracy sufficient to use it as a rough estimate even for quasi-2D systems within the parameter range studied here.

\subsection{The effect of interlayer coupling}\label{Sec:IL}

Critical temperatures are also affected by the interlayer interaction and this effect is not simple to describe. We have used atomistic simulations to obtain the interlayer interaction dependence for a range of $J_L$ that seems to be relevant for VI$_3$. Our model provides a reference for the finite-temperature behavior of a conventional quasi-two-dimensional ferromagnet and isolates the role of the interlayer exchange before local inhomogeneity is introduced.

\begin{figure}[H]
\centering
\begin{subfigure}{0.46\textwidth}
    \centering
    \includegraphics[width=\linewidth]{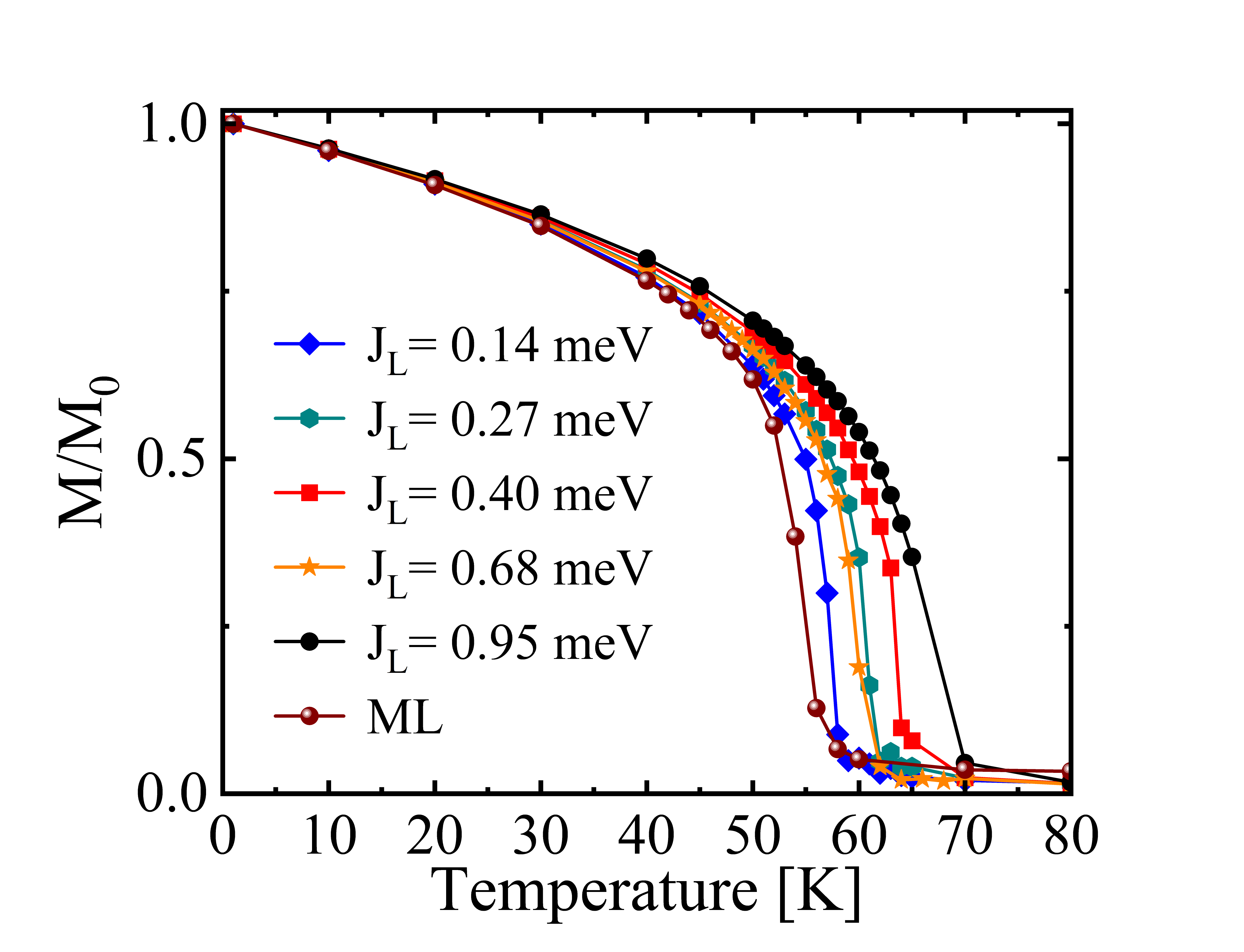}
    \caption{}
    \label{VI3_IL_M_Vs_T_SD_Bulk}
\end{subfigure}
\hfill
\begin{subfigure}{0.46\textwidth}
    \centering
    \includegraphics[width=\linewidth]{VI3_FC_SWT_Vs_homoASD.png}
    \caption{}
    \label{VI3_SD_HOMO_Irkhin_Comp_Bulk2}
\end{subfigure}
\caption{Magnetic properties of the homogeneous bulk model of \(\mathrm{VI}_3\). (a) Normalized magnetization \(M/M_0\) as a function of different effective interlayer couplings \(J_L\). (b) Curie temperature \(T_{\mathrm C}\) as a function of \(J_L\), obtained using UppASD and fluctuation-corrected self-consistent spin-wave theory (FC-SWT)\cite{Irkhin1999_Katsn_LayMagnetsThe} .}
\label{fig:interlayer_dependence}
\end{figure}

As shown in Fig.~\ref{fig:interlayer_dependence}(a), the magnetization curves shift systematically to higher temperature as \(J_L\) increases. The corresponding transition temperature rises nearly monotonically over the full range of couplings considered, as summarized in Fig.~\ref{fig:interlayer_dependence}(b). For the weakest interlayer coupling, \(J_L=0.14~\mathrm{meV}\), the bulk response approaches the monolayer limit. For the strongest coupling, \(J_L=0.95~\mathrm{meV}\), the transition is shifted to approximately \(70~\mathrm{K}\). This is the expected behavior of a quasi-two-dimensional ferromagnet: although the interlayer interaction is weak on the meV scale, it suppresses critical fluctuations and promotes coherence between neighboring layers. This sensitivity due to interlayer coupling is also consistent with the experimentally observed pressure dependence of \(T_{\mathrm C}\) in VI$_3$ \cite{Valenta2021_VI3_Pressure_Tc}.

We may ask whether the behavior can be explained with some analytical models. The mean-field approach (MF) predicts the linear dependence of $T_{\mathrm{C}}$ on the sum over all magnetic interactions with each site, $J_0$  \cite{r_06_tkdb_ExchInt_CurTemp_philmag}:

\begin{equation}
T_{\mathrm C}^{\mathrm{MF}}
=\frac{2J_0}{3k_B}.
\label{eq:mean_field_tc}
\end{equation}
If we consider only the nearest-neighbor intralayer interactions, in our model  $J_0=3J+J_L$ .  The contribution of the small  $J_L$ just sums up with the big intralayer one. From this we simply obtain the slope $\partial T_{C} / \partial J_{L} = 7.7 $ K/meV. The Mean-field is generally overestimated $T_{\mathrm C}$, but this slope is underestimated, which is apparent from the comparison with Fig.~\ref{fig:interlayer_dependence}(b).  The accuracy of this method is thus clearly worse for small $J_L$. The mean-field approach obviously cannot handle the Mermin-Wagner behavior in the limit of vanishing both $J_L$ and magnetic anisotropy, where the correlation length diverges exponentially as $T\rightarrow0$  \cite{Sengupta2003_qMC_varILInt_quasi2D}. Even in the quasi-2D case of relatively small $J_L$ mean-field appears to capture the dependence only partially. The big difference between these two couplings is not correctly taken into account by mean-field theory, which utilizes only an effective sum of all exchange interactions with neighbors. 

Studies of this problem within self-consistent spin-wave theory have led to a fluctuation-corrected self-consistent spin-wave-theory (FC-SWT) expression by Irkhin \textit{et al.}\cite{Irkhin1999_Katsn_LayMagnetsThe}. The transition temperature is then
obtained from
\begin{equation}
T_{\mathrm C}\simeq
\frac{4\pi J\bar S_0g_0}{
\ln\!\left[
\dfrac{2q_0^2}{D_c+2\alpha_c+\sqrt{D_c^2+4\alpha_c D_c}}
\right]
+2\ln\!\left(
\dfrac{4\pi J\bar S_0g_0}{T_{\mathrm C}}
\right)},
\label{eq:irkhin_tc}
\end{equation}
with
\begin{equation}
q_0=\sqrt{\frac{T_{\mathrm C}}{JS}},\qquad
\alpha_c=\frac{T_{\mathrm C}J_L}{4\pi(Jg_0)^2},\qquad
D_c=D_0\left(\frac{T_{\mathrm C}}{4\pi J\bar S_0g_0}\right)^2 .
\label{eq:irkhin_params}
\end{equation}
Here \(J\) is the nearest-neighbor intralayer exchange used in the spin-wave model, \(S\) is the spin magnitude, \(\bar S_0\) is the zero-temperature renormalized spin, \(g_0\) is the zero-temperature renormalization factor used in Ref.~\cite{Irkhin1999_Katsn_LayMagnetsThe}, and \(D_0\) is the bare anisotropy gap. The quantities \(q_0\), \(\alpha_c\), and \(D_c\) are, respectively, the thermal long-wavelength cutoff, the dimensionless interlayer coupling at \(T_{\mathrm C}\), and the renormalized anisotropy gap. They are determined self-consistently rather than introduced as independent fitting parameters.

The \(T_{\mathrm C}\)  obtained by solving Eqs.~\eqref{eq:irkhin_tc} and
\eqref{eq:irkhin_params}  is shown in Fig.~\ref{fig:interlayer_dependence}(b) as a function of interlayer coupling \(J_L\). The obtained values are closer to the experimental bulk value of about 50 K
\cite{Son2019_Bulk_VI3} than the mean-field result, but remain higher. This
discrepancy is not unexpected since the fluctuation-corrected spin-wave
expression assumes a homogeneous defect-free system and does not include
defect- or stacking-induced spatial variations of exchange interactions. The
comparison suggests that weak interlayer coupling alone is not sufficient to
explain the reduced bulk ordering temperature of \(\mathrm{VI}_3\).

\subsection{Model with two coexisting V configurations}
In this section we construct a minimal disorder model containing both V orbital configurations, whose coexistence has been indicated by neutron spectroscopy and x-ray absorption/x-ray magnetic circular dichroism measurements \cite{Lane2021_SO_excit_VI3,Hovancik2023_VI3_XMCD_OrbMom}. 
Note that Lane \textit{et al.} \cite{
Lane2021_SO_excit_VI3} and Hovan\v{c}\'ik \textit{et al.} \cite{Hovancik2023_VI3_XMCD_OrbMom}  have used the term domains, which can evoke spatial segregation of the two types. However, the model used to fit the magnon spectra does not contain any segregation, and in fact signatures of disorder are encountered  \cite{Lane2021_SO_excit_VI3}. Therefore modeling the problem as a random mixture of two V types appears to be appropriate.

\begin{table}[H]
\centering
\caption{Magnetic parameters used in the two-environment spin-Hamiltonian model of VI$_3$. HO and LO denote high- and low-orbital-moment V environments, respectively. Here \(J_1\) is the nearest-neighbor intralayer exchange, \(J_L\) is the effective interlayer exchange per V site, and \(K\) is the single-ion anisotropy. The final column specifies the origin of each parameter.}
\label{heisenberg_parameters}
\begin{tabular}{lcccc}
\hline
Environment & \(K\) (meV) & \(J_1\) (meV) & \(J_L\) (meV) & Source\\
\hline
HO & 15.92 & 3.03 & 0.54 &
\(\,K,J_1\): Ref.~\cite{Yang2020_VI3_2DIsing_CFdiag}; \(J_L\): present DFT\\
LO & 0.40 & 2.18 & -0.38 &
\(\,K\): Ref.~\cite{Wang2020_VI3_magFPcalc}; \(J_1\): Ref.~\cite{Gordon2021_MagInt_VI3}; \(J_L\): present DFT\\
\hline
\end{tabular}
\end{table}

The parameters entering Eq.~\ref{hamiltonian} in our model are listed in Table~\ref{heisenberg_parameters}. For the SIA of V we adopt the published disparate values obtained by ab initio calculations, 15.9 meV for the HO type \cite{Yang2020_VI3_2DIsing_CFdiag} and 0.4 meV (upper limit) for the LO type \cite{Wang2020_VI3_magFPcalc}. The three nearest-neighbor intralayer interactions for the LO case were taken from calculations by Gordon et al. \cite{Gordon2021_MagInt_VI3}. Notably, the 2\textsuperscript{nd} and 3\textsuperscript{rd}  nearest-neighbor interactions appear to be small in VI$_3$ and do not play a significant role compared to other effects considered here. For the HO configuration, we employ intralayer interactions calculated by comparing the energies of the FM and Neel AFM state \cite{Yang2020_VI3_2DIsing_CFdiag}. See Tab. \ref{heisenberg_parameters} for a summary of the parameters used.
We are unaware of any published calculation of the interlayer interaction for bulk  VI$_3$ in the HO state. The interlayer exchange is mediated by extended V--I$\cdots$I--V super-superexchange channels across the van der Waals gap \cite{Jiang2021_2Dmag_Review} and is therefore relatively small and highly sensitive to changes in the interlayer geometry and stacking registry; different stackings can even reverse its sign \cite{Gordon2021_MagInt_VI3}. Although direct experimental confirmation of this effect in VI$_3$ is still lacking, analogous sensitivity has been demonstrated in other vdW trihalides \cite{Song2019_Switch2DMag_Pressure_CrI3,Li2019_Pressure_Stacking_IntLExch,Chen2019_CrBr3_StackDep_Interlay,Misek2025_CrBr3}. This makes a qualified estimate of its value rather challenging. We have thus performed first principles calculations of these interactions for both configurations, as detailed in Methods section. The values obtained (Tab. \ref{heisenberg_parameters}) highlight its sensitivity also to the electronic configuration, as the sign also differs for the two cases considered. An interesting point is that in both our calculations and other published calculations \cite{Gordon2021_MagInt_VI3,Wang2020_VI3_magFPcalc} for a fixed geometry the interlayer interactions are, in fact, not too small compared to intralayer interactions $\left|J_{L}\right|>J_{1}/10$. However, the fact that it may change sign upon small distortion means that in the presence of polarons there will be a large cancelation on average, and this key feature is also captured by our model with two V types. Simulations with polarons would probably be able to describe this situation quantitatively even more accurately, but these are extremely numerically demanding. Therefore, we leave it for the future as a potential step in improving our model.

\begin{figure}[H]
\centering
\begin{subfigure}{0.46\textwidth}
\centering
\includegraphics[width=\linewidth]{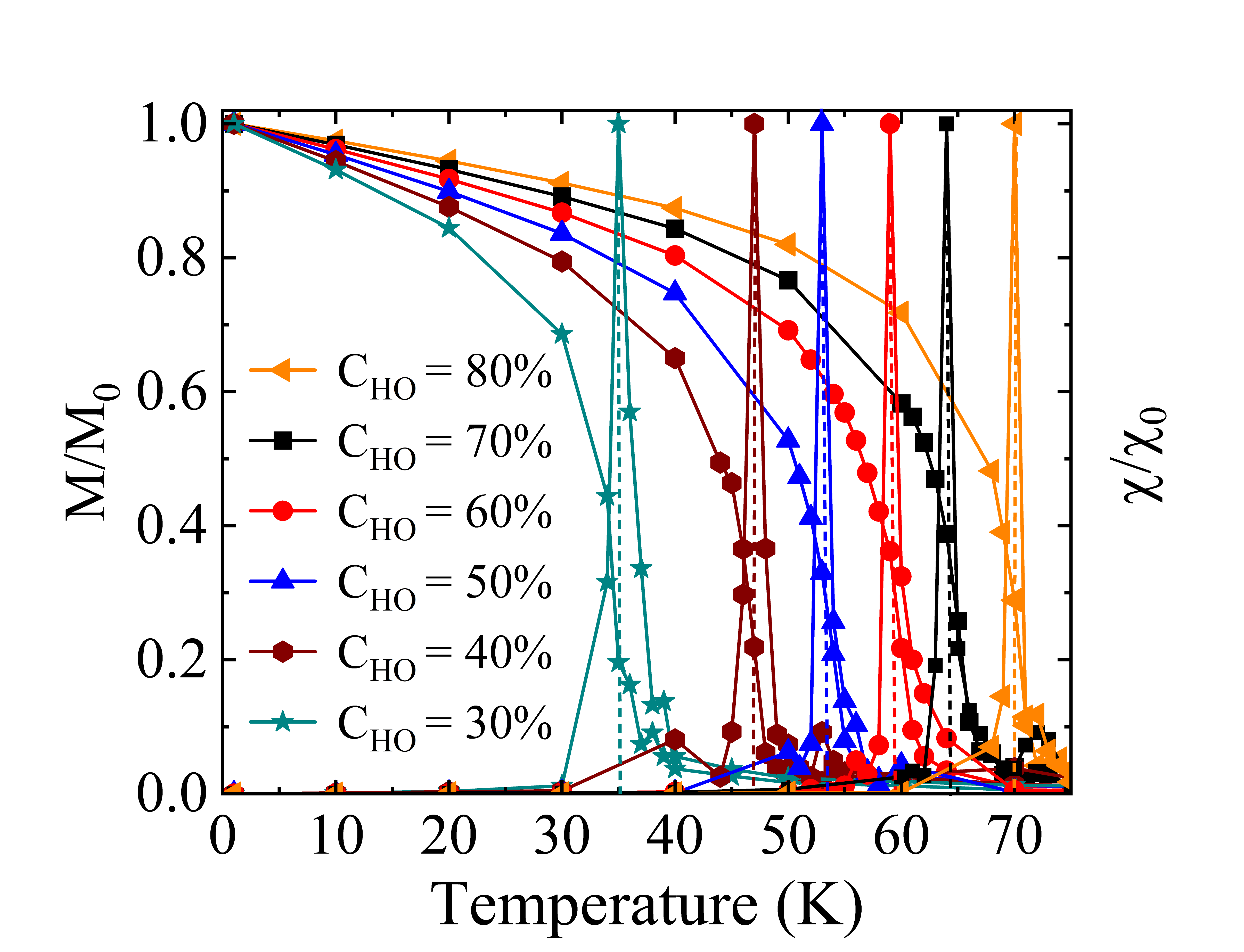}
\caption{}
\label{fig:two_domain_mt_c}
\end{subfigure}
\hfill
\begin{subfigure}{0.46\textwidth}
\centering
\includegraphics[width=\linewidth]{VI3_quasi2D_Vs_ML_SetBGordon_bulk_ml_comp.png}
\caption{}
\label{fig:two_domain_tc_c}
\end{subfigure}
\caption{
Effect of the HO/LO concentration in the two-environment (HO+LO) bulk model of $\mathrm{VI}_3$.
(a) Temperature dependence of the normalized magnetization, $M/M_0$, and susceptibility, $\chi/\chi_0$, for different  $C_{\mathrm{HO}}$.
(b) Curie temperature $T_{\mathrm{C}}$ of bulk and monolayer as a function of the ${\mathrm{HO}}$ concentration , $C_{\mathrm{HO}}$.
}
\label{fig:two_domain_concentration}
\end{figure}

The methods used in the key studies  \cite{Lane2021_SO_excit_VI3,Hovancik2023_VI3_XMCD_OrbMom} do not allow for a precise determination of the HO/LO type concentration. Its authors have assumed, in both cases, a concentration of 50$\%$, which has allowed the reproduction of observations with sufficient precision. Therefore, we also show the results for the case of $C_{\mathrm{HO}}=0.5$, and then evaluate $T_{\mathrm{C}}$ for a reasonable range of $C_{\mathrm{HO}}$ between 0.2 and 0.8. In Fig.~\ref{fig:two_domain_concentration}(a), the temperature-dependent normalized magnetization and susceptibility are shown for the representative mixed configuration with various concentrations $C_{\mathrm{HO}}$. The magnetization decreases rapidly in the transition region, whereas the susceptibility shows a clear peak at the ordering temperature. 

The dependence of $T_{\mathrm{C}}$ on the relative concentration of the two V environments for both bulk and ML  $\mathrm{VI}_3$ is summarized in Fig.~\ref{fig:two_domain_concentration}(b). Increasing $C_{\mathrm{HO}}$ systematically increases the bulk ordering temperature, from approximately $24~\mathrm{K}$ in $C_{\mathrm{HO}}=20\%$ to approximately $70~\mathrm{K}$ in $C_{\mathrm{HO}}=80\%$ . Changing $C_{\mathrm{HO}}$ also changes the number and connectivity of the exchange paths between layers HO--HO, LO--LO, and HO--LO. Since these paths can carry different exchange strengths and, in some cases, different signs, the interlayer exchange field becomes spatially non-uniform. The resulting concentration dependence $T_{\mathrm{C}}(C_{\mathrm{HO}})$ here reflects the competition between robust intralayer ferromagnetic correlations and frustrated interlayer coupling.
Comparing the bulk $T_{\mathrm{C}}$ with the prediction allows us to determine $C_\mathrm{HO}^{\mathrm{bulk}}=$0.46 within the assumptions of our model. In order to connect this concentration to the presence of polarons, one can assume that the distortion around a trapped electron encompasses V up to the 2nd nearest neighbors, and hence 10 V atoms are affected  \cite{Mastrippolito2023_VCl3_Polaronic_MottInt}. To achieve a similar amount of both affected and unaffected V atoms, roughly 1 electron per 20 formula unit (f.u) is needed, which corresponds to the doping of 0.0125 electron per atom. Such doping appears to be plausible in real $\mathrm{VI}_3$ samples.

To separate the effect of anisotropy contrast from that of exchange inhomogeneity, we construct another model where we use a fictitious V configuration MLH (mixed LO and HO). Its exchange parameters are equal to those of the HO configuration, but its anisotropy is equal to the LO configuration $(J^{\mathrm{MLH}}=J^{\mathrm{HO}},\ K_{\mathrm{MLH}}= K_{\mathrm{LO}})$. When we embed these in an alloy together with HO type atoms, only the anisotropy between these atoms differs. Thus, in the HO+MLH model, there is no exchange inhomogeneity, unlike the exchange dependent HO+LO model.

\begin{figure}[H]
\centering
\begin{subfigure}{0.46\textwidth}
\centering
\includegraphics[width=\linewidth]{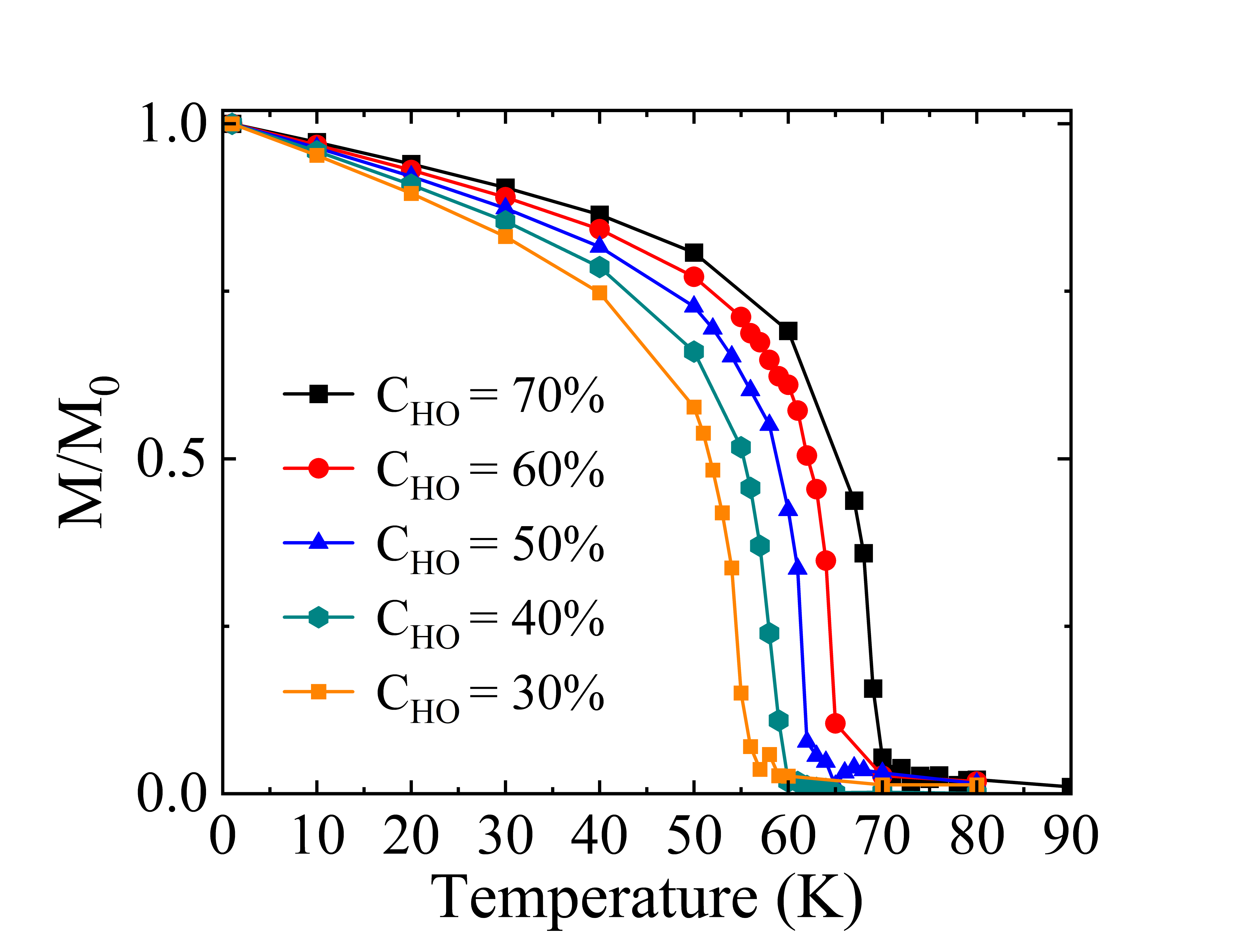}
\caption{}
\label{M_T_diff_K_only}
\end{subfigure}
\hfill
\begin{subfigure}{0.46\textwidth}
\centering
\includegraphics[width=\linewidth]{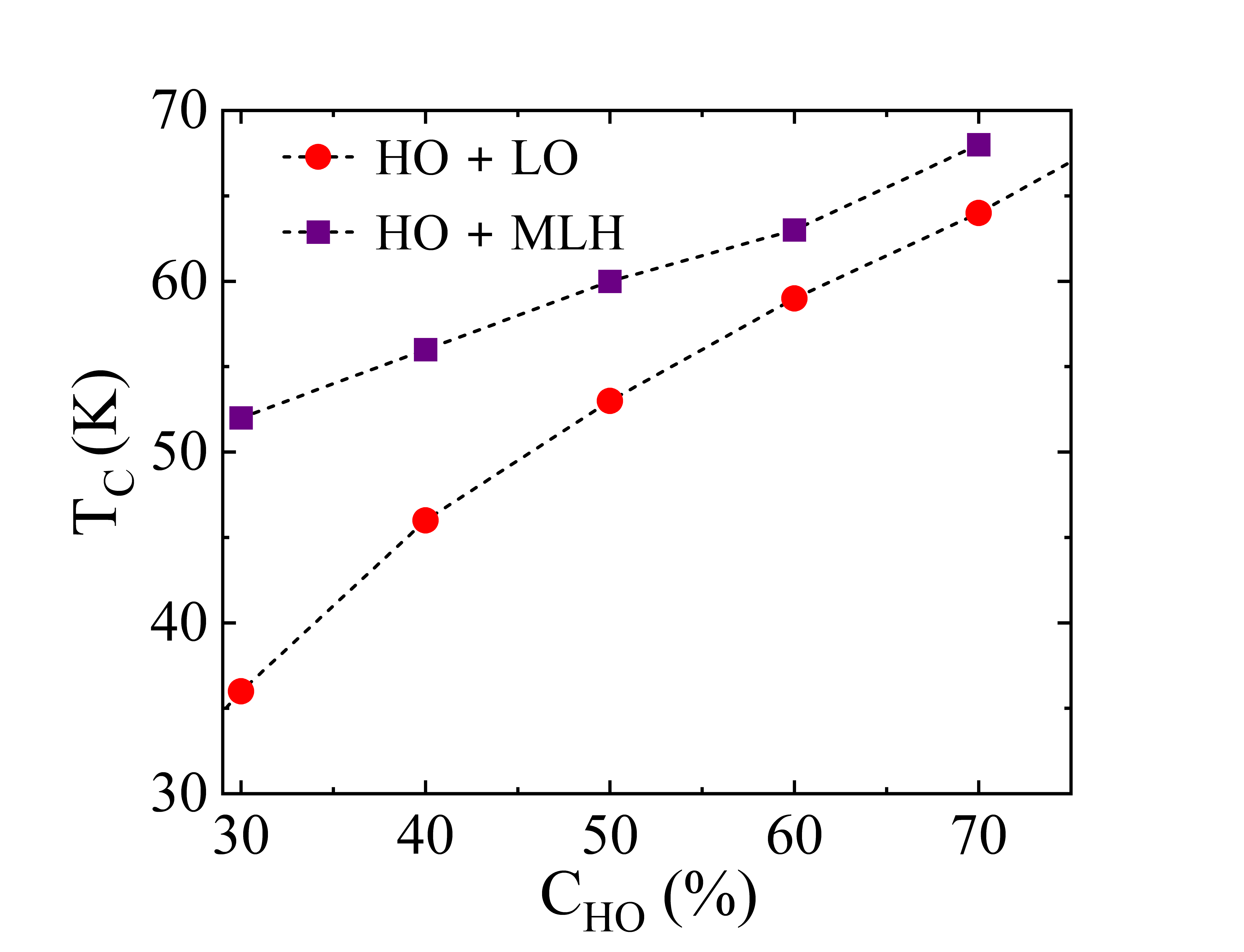}
\caption{}
\label{VI3_different_J_comparision_Bulk}
\end{subfigure}
\caption{
Effect of anisotropy contrast and exchange inhomogeneity in the mixed HO/LO bulk model of $\mathrm{VI}_3$.
(a) Temperature-dependent normalized magnetization, $M/M_0$, for different HO concentrations in the control model with environment-independent exchange parameters, and environment-dependent anisotropy, HO+MLH atoms only
(b) Curie temperature $T_{\mathrm{C}}$ as a function of $C_\mathrm{HO}$ for the anisotropy-only control model, HO+MLH type mixture, and for the full two-environment model, HO+LO type mixture (the same as in Fig.~\ref{fig:two_domain_concentration})
}
\label{fig_diff_K_only}
\end{figure}

Figure~\ref{fig_diff_K_only} shows the temperature-dependent normalized magnetization for the anisotropy-only control model (HO+MLH). As $C_{\mathrm{HO}}$ decreases, the magnetization drop shifts to a lower temperature. This behavior is expected because the fraction of MLH sites increases, and MLH sites have a single-ion anisotropy much smaller than that of HO sites. The reduced anisotropy lowers the energy cost of transverse spin fluctuations and weakens the stability of the out-of-plane ordered state. Thus, even when the exchange network is kept homogeneous, replacing HO sites with MLH sites is sufficient to reduce the Curie temperature, as seen in  Fig.~\ref{fig_diff_K_only}(b).  The relatively moderate slope of the HO+MLH curve indicates that anisotropy contrast alone significantly changes the transition temperature, but exchange-related effects are also important.

When the exchange parameters are also allowed to depend on the atomic type, the original HO+LO model, the Curie temperature decreases more strongly, particularly in the LO-rich region (shown in both \ref{fig:two_domain_concentration}(b) and Fig. ~\ref{fig_diff_K_only}(b)). This indicates that exchange inhomogeneity provides an additional mechanism for reducing the thermal stability of the bulk magnetic state. The effect is most visible at low $C_{\mathrm{HO}}$, where the difference between the two curves is the largest. On the other hand, at high $C_{\mathrm{HO}}$ the two models approach nearly the same transition temperature. This indicates that the magnetic order of the system becomes dominated by HO sites in this regime, and LO sites cannot achieve extended fluctuations on their own sublattice.
These results show that the reduction of $T_{\mathrm{C}}$ with decreasing $C_{\mathrm{HO}}$ is not controlled by anisotropy averaging alone. The smaller anisotropy of LO sites weakens the local stabilization of the ordered moments, whereas the environment-dependent exchange modifies the connectivity of the HO--HO, LO--LO, and HO--LO exchange paths. As a result, the effective magnetic coupling becomes spatially non-uniform, further suppressing the bulk ordering temperature compared to that of the anisotropy-only model.

\begin{figure}[H]
\centering
\includegraphics[width=0.8\linewidth]{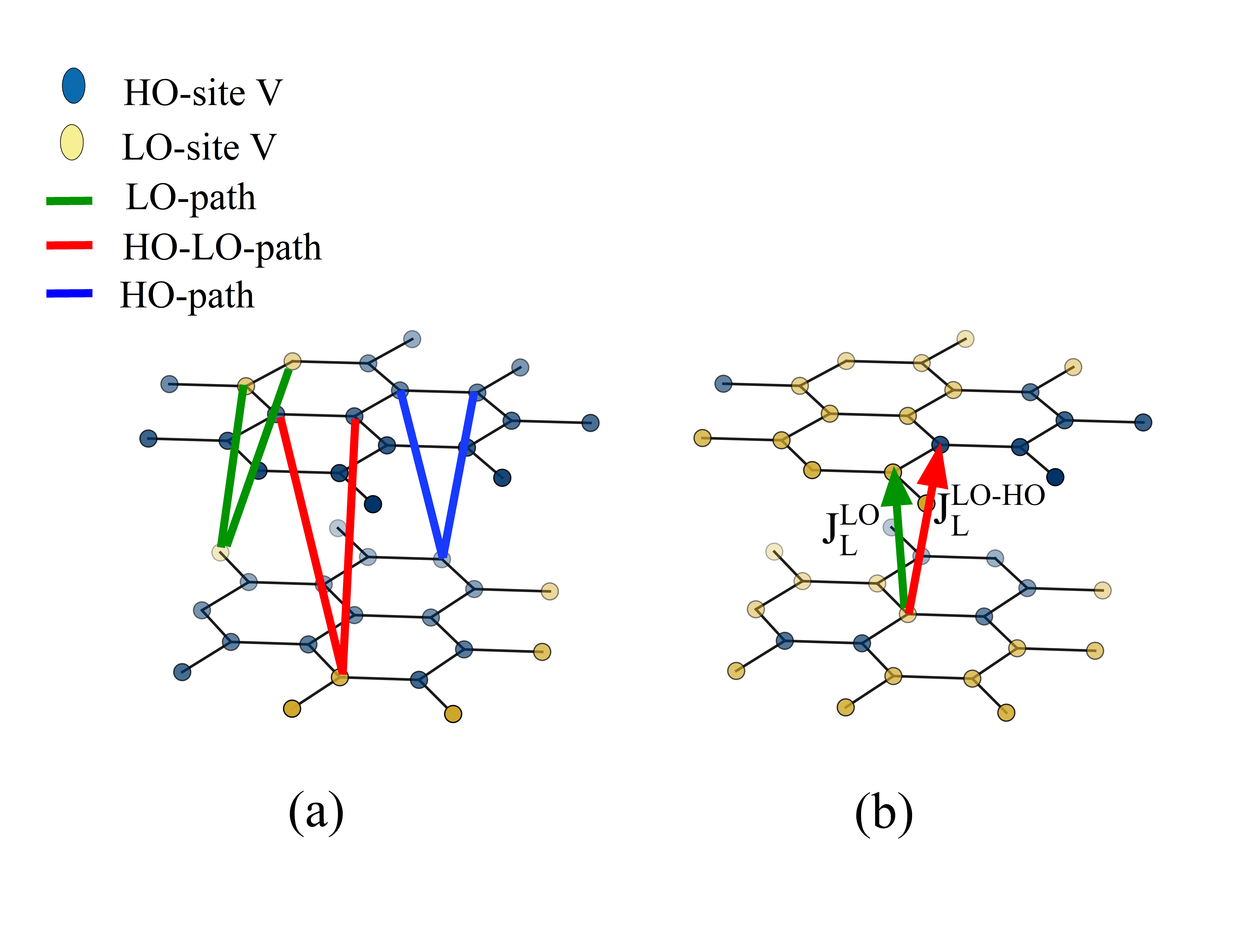}
\caption{
Schematic representation of interlayer V--I$\cdots$I--V super-superexchange paths in the mixed HO/LO network of rhombohedral $\mathrm{VI}_3$.
Blue and yellow spheres denote high- and low-orbital-momentum V sites, respectively.
blue, green and red lines indicate HO--HO, LO--LO, and HO--LO interlayer paths.
Panels (a) and (b) show representative configurations with $C_{\mathrm{HO}}=70\%$ and $C_{\mathrm{HO}}=30\%$, respectively.
The arrows in panel (b) highlight competing interlayer exchange channels, such as $J_L^{\mathrm{LO}}$ and $J_L^{\mathrm{LO-HO}}$.
}
\label{fig:HO_LO_interlayer_paths}
\end{figure}

The effect of interlayer exchange variation on the concentration-dependent $T_{\mathrm{C}}$ is important for understanding the difference between the bulk and ML cases.  When two local orbital environments coexist,  interlayer exchange channels are no longer equivalent: like-site and mixed-site paths can differ in both strength and sign. The bulk interlayer coupling therefore forms a spatially non-uniform exchange network rather than a single homogeneous field. Figure~\ref{fig:HO_LO_interlayer_paths} shows that changing the concentration does not simply replace one type of V site with another; it reorganizes the topology of the interlayer exchange network. In the high-orbital-rich limit, the coupling between adjacent layers is carried out mainly by similar local environments. As the fraction of low-orbital sites increases, low-orbital and mixed interlayer paths become more frequent. Because these V--I$\cdots$I--V geometries involve different orbital characters, they can contribute differently to the effective interlayer exchange and introduce competing magnetic channels.

This provides a microscopic route for suppressing the bulk ordering temperature. In the homogeneous model, the presence of $J_L$ always enhances $T_{\mathrm{C}}$. However, in the mixed-orbital network, the spatial variation of $J'_{ij}$ reduces the coherence of the interlayer exchange field while preserving strong intralayer ferromagnetic correlations. Note that the difference between bulk and ML $T_{\mathrm{C}}$ (shown in Fig.  \ref{fig:two_domain_concentration}(b)) decreases with decreasing $C_\mathrm{HO}$, and therefore the stabilization due to intralayer exchange is less relevant as the competition between LO and HO type interaction becomes more balanced. For $C_\mathrm{HO}<0.4$ the ML $T_{\mathrm{C}}$ is even becoming higher than that of bulk. 

\begin{figure}[H]
\centering
\includegraphics[width=0.6\linewidth]{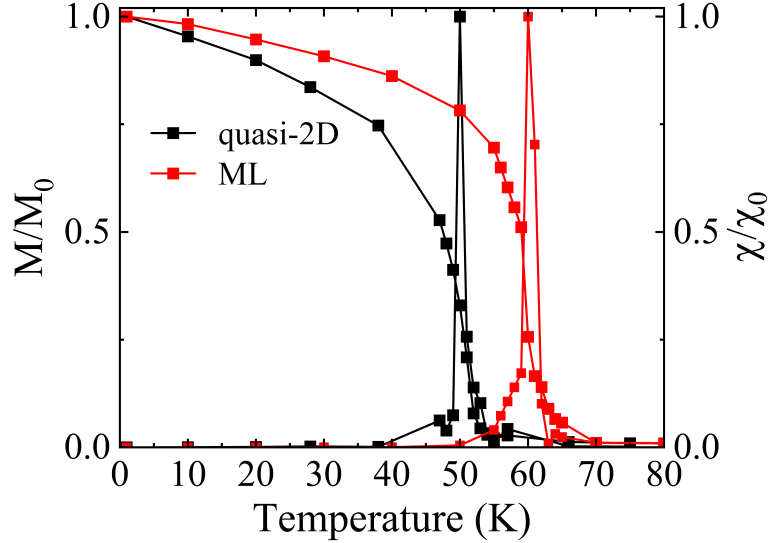}
\caption{Temperature dependence of the normalized magnetization, $M/M_0$, and susceptibility, $\chi/\chi_0$, for the HO/LO mixed model of bulk and monolayer VI$_3$; each of them assigned its own separate predicted HO atom concentrations, $C_\mathrm{HO}^{\mathrm{bulk}}$ and $C_\mathrm{HO}^\mathrm{ML}$  }

\label{fig:two_domaain_diff_CHO_Bulk_monolayer_SetBGordon}
\end{figure}

The exfoliation used to obtain an ML \(\mathrm{VI}_3\) \cite{Lin2021_VI3_MCD_ML} can lead to a change of the possible excess charge on the V atoms. Therefore, also the concentration of polarons and the concentration  $C_{HO}$ may differ from the bulk value. Agreement with the experimentally observed $T_{\mathrm{C}}=$ 60K for an mono-layer is achieved for $C_\mathrm{HO}^\mathrm{ML}$ approximately $0.76$. Thus, the combined change of single-ion anisotropy and exchange interactions between the two possible V configurations allows us to propose parameters that describe both bulk and ML VI$_3$ behavior  (shown in Fig.~\ref{fig:two_domaain_diff_CHO_Bulk_monolayer_SetBGordon}). Note that a comparison of surface and bulk-sensitive spectroscopies has revealed that occupation of the ${a_{1g}}$ level is more pronounced on surfaces \cite{Vita2022_OrbChar_VI3_CrI3_GS_ARPES, DeVita2026_VI3_ValenMod_XRay}, again indicating volatility of the HO/LO concentration and probably higher concentration of the \(\mathrm{VI}_3\) HO state when not in bulk.

\section{Conclusions}

We have used atomistic spin-dynamics simulations for a layered spin Hamiltonian to understand the behavior of critical temperatures in both bulk and ML \(\mathrm{VI}_3\). 
We have first examined the role of single-ion anisotropy and interlayer exchange interactions,  phenomena that stabilize magnetic order in quasi-two-dimensional magnets. 
We have performed calculations for homogeneous systems with parameter ranges that are known to be achievable due to the possibility of two V atom configurations. 
The results show to what extent the critical temperature can be modified by single-site anisotropy and interlayer exchange. In particular, the variation with anisotropy allows us to expect a large change in $\mathrm{T_C}$ when the occupation of different types of V changes.

The subsequent simulations model approximately the coexistence of both atom types. This coexistence represents a highly complex problem, where the components respond differently to thermal fluctuations, mainly due to their different anisotropy. The geometry of V--I$\cdots$I--V interlayer super-superexchange is also strongly perturbed and leads to competing interlayer exchange pathways. The interlayer exchange network becomes spatially non-uniform and frustrated. This frustration weakens the global magnetic coherence of the bulk crystal and suppresses the ordering temperature. For this mixture, the presence of interlayer interaction does not significantly increase the critical temperature compared to the monolayer case, where competing interlayer pathways are absent. Our findings emphasize that weak interlayer interactions in van der Waals magnets should not always be viewed as simple stabilizing perturbations; when they are spatially non-uniform or competing, they can instead suppress magnetic order. This insight may be important for understanding and engineering thickness-dependent magnetism in VI$_3$ and related layered magnetic materials.

 Our results show that variation of the ratio between different V types can cause a rather large change in $T_{\mathrm{C}}$. This behavior highlights the sensitivity of VI$_3$ magnetism to local orbital and structural conditions. Our simulations predict that the bulk system has  $T_{\mathrm{C}}$ close to the experimental one for $C_\mathrm{HO}^{\mathrm{bulk}}=46\%$.  This finding is in good agreement with other experiments, where the concentration of 50\% allowed the observations to reproduce. Thus, we have provided further evidence for the picture of two coexisting V configurations in $\mathrm{VI}_3$, this time on the basis of finite-temperature magnetization behavior.
  We have found that the different critical temperatures of the bulk and ML samples could be explained by a variation of the concentration of V atom types with different electronic configurations. Among the different properties of these two types, the variation of anisotropy plays a key role in finite-temperature magnetization. Our results further suggest that controlling the relative occupation of the HO and LO V configurations could tune the Curie temperature $T_{\mathrm{C}}$ over a wide range, with at least a factor of two differences between its upper and lower limits.

\section*{Acknowledgements}
This work is part of the research project GA Č R 25-15448S, funded by the Czech Science Foundation, and the project GAUK 128624, funded by the Charles University Grant Agency.
This work was supported by the Ministry of Education, Youth and Sports of the Czech Republic, project Quantum materials for applications in sustainable technologies (QM4ST), funded as project no. CZ.02.01.01/00/22\_008/0004572 by P JAK, call Excellent Research, and project e-INFRA CZ (ID:90254).

{\footnotesize{}\printbibliography
}{\footnotesize\par}
\end{document}